\documentstyle[12pt]{article}

\catcode`\@=11
\def\marginnote#1{}
\newcount\hour
\newcount\minute
\newtoks\amorpm
\hour=\time\divide\hour by60
\minute=\time{\multiply\hour by60 \global\advance\minute
by-\hour}\edef\standardtime{{\ifnum\hour<12
\global\amorpm={am}%
        \else\global\amorpm={pm}\advance\hour by-12 \fi
        \ifnum\hour=0 \hour=12 \fi
        \number\hour:\ifnum\minute<10
0\fi\number\minute\the\amorpm}}
\edef\militarytime{\number\hour:\ifnum\minute<10
0\fi\number\minute}

\def\draftlabel#1{{\@bsphack\if@filesw {\let\thepage\relax
   \xdef\@gtempa{\write\@auxout{\string
      \newlabel{#1}{{\@currentlabel}{\thepage}}}}}\@gtempa
   \if@nobreak \ifvmode\nobreak\fi\fi\fi\@esphack}
        \gdef\@eqnlabel{#1}}
\def\@eqnlabel{}
\def\@vacuum{}
\def\draftmarginnote#1{\marginpar{\raggedright\scriptsize\tt#1}}
\def\draft{\oddsidemargin -.5truein
        \def\@oddfoot{\sl preliminary draft \hfil
        \rm\thepage\hfil\sl\today\quad\militarytime}
        \let\@evenfoot\@oddfoot \overfullrule 3pt
        \let\label=\draftlabel
        \let\marginnote=\draftmarginnote

\def\@eqnnum{(\theequation)\rlap{\kern\marginparsep\tt\@eqnlabel}%
\global\let\@eqnlabel\@vacuum}  }

\def\numberbysection{\@addtoreset{equation}{section}
        \def\theequation{\thesection.\arabic{equation}}}

\def\underline#1{\relax\ifmmode\@@underline#1\else
 $\@@underline{\hbox{#1}}$\relax\fi}

\catcode`@=12
\relax

\numberbysection

\advance \topmargin by -\headheight
\advance \topmargin by -\headsep

\evensidemargin \oddsidemargin
\def\rf#1{(\ref{#1})}
\def\lab#1{\label{#1}}
\def\nonu{\nonumber}
\def\br{\begin{eqnarray}}
\def\er{\end{eqnarray}}
\def\be{\begin{equation}}
\def\ee{\end{equation}}

\def\({\left(}
\def\){\right)}

\newcommand{\ct}[1]{\cite{#1}}
\newcommand{\bi}[1]{\bibitem{#1}}

\relax

\def\b{\beta}

\def\h{{1\over 2}}

\def\pa{\partial}

\def\tp0{\Theta_{+}^{(0)}}
\def\tm0{\Theta_{-}^{(0)}}

\def\f#1#2#3 {f^{#1#2}_{#3}}

\def\win1{{\sf w_{1+\infty}}}

\def\Win1{{\sf W_{1+\infty}}}

\def\rlx{\relax\leavevmode}
\def\inbar{\vrule height1.5ex width.4pt depth0pt}
\def\IZ{\rlx\hbox{\sf Z\kern-.4em Z}}
\def\IR{\rlx\hbox{\rm I\kern-.18em R}}
\def\IC{\rlx\hbox{\,$\inbar\kern-.3em{\rm C}$}}
\def\IN{\rlx\hbox{\rm I\kern-.18em N}}
\def\IO{\rlx\hbox{\,$\inbar\kern-.3em{\rm O}$}}
\def\IP{\rlx\hbox{\rm I\kern-.18em P}}
\def\IQ{\rlx\hbox{\,$\inbar\kern-.3em{\rm Q}$}}
\def\IF{\rlx\hbox{\rm I\kern-.18em F}}
\def\IG{\rlx\hbox{\,$\inbar\kern-.3em{\rm G}$}}
\def\IH{\rlx\hbox{\rm I\kern-.18em H}}
\def\II{\rlx\hbox{\rm I\kern-.18em I}}
\def\IK{\rlx\hbox{\rm I\kern-.18em K}}
\def\IL{\rlx\hbox{\rm I\kern-.18em L}}
\def\one{\hbox{{1}\kern-.25em\hbox{l}}}
\def\0#1{\relax\ifmmode\mathaccent"7017{#1}%
B        \else\accent23#1\relax\fi}

\def\PRL#1#2#3{{\sl Phys. Rev. Lett.} {\bf#1} (#2) #3}

\def\PRD#1#2#3{{\sl Phys. Rev.} {\bf D#1} (#2) #3}
\def\PRE#1#2#3{{\sl Phys. Rev.} {\bf E#1} (#2) #3}
\def\PRv#1#2#3{{\sl Phys. Rev.} {\bf #1} (#2) #3}

\def\JMP#1#2#3{{\sl J. Math. Phys.} {\bf #1} (#2) #3}

\def\SPTP#1#2#3{{\sl Suppl. Prog. Theor. Phys.} {\bf #1} (#2) #3}

\begin{document}

\topmargin -.6in

\begin{titlepage}

May, 1999 \hfill{}
\vskip .6in

\begin{center}
{\large {\bf Relativistic Quantum Thermodynamics of Ideal Gases in $2$ Dimensions}}
\end{center}

\normalsize
\vskip .4in

\begin{center}

H. Blas and B. M. Pimentel 

\par \vskip .1in \noindent

Instituto de F\'{\i}sica Te\'{o}rica-UNESP\\
Rua Pamplona 145\\
01405 S\~{a}o Paulo, Brazil\\
\par \vskip .1in \noindent
and
\par \vskip .1in \noindent
J. L. Tomazelli
\par \vskip .1in \noindent
DFQ-Faculdade de Engenharia, Universidade Estadual Paulista-\\
Campus de Guaratinguet\'a, Av. Dr. Ariberto Pereira da Cunha, 333,\\
12500-000 Guaratinguet\'a, S.P., Brazil\\
E-mail: pimentel@ift.unesp.br
\par \vskip .2in
\end{center}

\begin{center}
{\large {\bf ABSTRACT}}\\
\par \vskip .4in
\end{center}
In this work we study the behavior of relativistic ideal Bose and Fermi gases in two space dimensions. Making use of polylogarithm functions we derive a closed and unified expression for their densities. It is shown that both type of gases are essentially inequivalent, and only in the non-relativistic limit the spinless and equal mass Bose and Fermi gases are equivalent as known in the literature.   

\par \vskip .3in \noindent

\end{titlepage}

\section{Introduction}
The polylogarithm and hyperlogarithm functions \ct {lewin, kirilov} have many intriguing properties, with important applications in many branches of mathematics and physics such as number theory, representation theory of infinite dimensional algebras, exact solvable models,  conformal field theory, etc. The remarkable dilogarithm will be the main protagonist of our paper.
Recently, in the context of statistical thermodynamics of ideal gases at low temperature, in any dimension, it has been observed that the reduced density can be expressed in polylogarithms of integral and half-integral order \ct{haber1,haber}. Polylogarithms turned out to give a single unified picture of the density of the ideal Fermi and Bose gases and also the classical gas. Different statistical effects are related to different structural properties of polylogarithms. In this way, it has been stablished that different statistics represent different domains of polylogarithms \ct{lee1}.
 
In particular, it is interesting to consider $2$ dimensional ideal gases at low temperature. It was shown by May \ct{may} that $2D$ ideal Bose and Fermi gases have the same specific heat at the same temperature. Making use of some remarkable properties of the polilogs, M.H. Lee has shown that the $2D$ ideal nonrelativistic Bose and Fermi gases are completely equivalent (see \ct{lee1,lee2} and references therein).

The aim of this paper is to consider a relativistic dispersion relation and show that the results of \ct{lee2} can be obtained as a non-relativistic limit of the relativistic ideal Fermi and Bose gases. We shall show that in $2$ dimensions the statistical thermodynamics of ideal quantum gases can be unified even for a relativistic dispersion relation and an equivalence between the Bose and Fermi gases stablished for the non-relativistic limit. This unified formulation is stablished through a formulation of the statistical thermodynamics of ideal gases by polylogs. 

\section{The partition function} 

The functional integral expression for the partition function at finite temperature can be formulated directly in real time as well as in imaginary time formalism. In this work we adopt the latter one. To study the thermodynamics of relativistic bosons (fermions) at finite charge density and temperature in the grand canonical ensemble, one has to compute the grand partition function 
\br
\lab{zeta1}
Z={\rm Tr} e^{-\b(H-\mu \hat{N})}=\int d\Phi_{a}<\Phi_{a}|e^{-\b(H-\mu\hat{N})}|\Phi_{a}>,
\er
where the sum runs over all states and $\mu$ is a chemical potential associated to some conserved charge of the system. This partition function can be written as \ct{ber}
\br
\lab{zeta2}
Z=\int[d\Pi]\int_{\begin{array}{ll}{\rm periodic}\\
({\rm antiperiodic})\end{array}}[d\Phi]exp\left\{\int_{0}^{\b}d\tau\int d^{3}x \left(i\Pi \frac{\pa \Phi}{\pa \tau}-{\cal H}(\Pi,\Phi)+\mu {\cal N}(\Pi,\Phi)\right)\right\},
\er 
where $\tau$ is real Euclidean time. The term periodic (antiperiodic) means that the integration over the field is constrained so that $\Phi ({\bf x},0)=\pm \Phi ({\bf x},\b)$, where the upper sign $(+)$ refers to bosons and the lower sign $(-)$ refers to fermionic fields. This is a consequence of the trace operation.
Since all other standard thermodynamic properties may be determined from $Z$ we must compute this function. If we consider a theory of a charged scalar free field for the boson Lagrangian and the theory of a free fermion (Dirac) Lagrangian, the actions are quadratic in their corresponding fields; then all the integrations in \rf{zeta2} are Gaussian and can be performed exactly (see \ct{kapusta} and references therein). The outcome of these computations is
\be
\lab{logz}
{\rm ln} Z = \mp V \int \frac{d^3k}{(2\pi)^3}\left[\b\omega + {\rm ln}(1\mp e^{-\b(\omega-\mu)})+{\rm ln}(1\mp e^{-\b(\omega+\mu)}) \right],
\ee
where $\omega = (k^{2}+m^{2})^{1/2}$ and $\beta =T^{-1} $ (in units of $\hbar=c=1$). 

There are several observations to make about \rf{logz}. An overall spin factor of $2$ must be written on the right hand side in the fermion case (in 3D spatial dimension), $V$ is the volume of $x$ space, the upper signs hold for bosons and the lower ones for fermions . Separate contributions from particles ($\mu$) and antiparticles ($-\mu$) are evident. It must also be noticed that the zero-momentum mode contribution, in the boson case, has not been considered. Finally, the zero-point energies of the respective vacua also appear in this formula. 

Since we are interested in low temperature (temperatures smaller than the masses of the particles) thermodynamic quantities, we will neglect the third term (antiparticle contribution) of \rf{logz}. In the grand canonical ensemble, the expression for the reduced density $\rho$ ( $\rho=N/V$) is obtained from $Z$ by the relation, $\rho=\frac{\b^{-1}}{V} \frac{\pa {\rm ln}Z}{\pa \mu}$, then one gets
\br
\lab{stat}
\rho &=& \int \frac{d^3k}{(2\pi)^3}\frac{1}{\exp[\beta(\omega_{k}-\mu)]\mp1}.
\er
As usual the upper (lower) sign in \rf{stat} correspond to the Bose (Fermi) case. In the case of bosons we must require $\mu\leq m$ in order to ensure a positive-definite value for $n_{k}$, the number of bosons with momentum $k$.

\section{Statistical thermodynamics of two dimensional ideal relativistic quantum gases: unified formulation}

The reduced density of an ideal gas in $d$-spatial dimensions is expressible as \ct{haber1,haber}
\br
\lab{densi}
\rho &=& \pi^{-\frac{d+1}{2}}\,\Gamma(\frac{d+1}{2})\,T^{d}\,g_{d}(\overline{m}, r),
\er
where we have difined the dimensionless variables $\overline{m}=m/T$ and $r=\mu/m$ (note that $|r|\leq 1$ in the case of bosons) and the function
\br
\lab{func}
g_{d}(\overline{m},r) &=& \frac{1}{\Gamma(d)}\int_{0}^{+\infty}dx\, x^{d-1}\frac{1}{\exp[(x^{2}+\overline{m}^{2})^{1/2}-r\overline{m}]\mp1}.
\er 

A method for obtaining low-temperature expansions in term of polylogs and for any dimension $d$ is presented in references \ct{haber}. Then, the expression for the function $g_{d}(\overline{m},r)$ can be written as

\br
\lab{sum}
g_{d}(\overline{m},r)&=&(2\overline{m})^{d/2-1}\frac{\sqrt{\pi}}{2^{d-1}\Gamma(\frac{d+1}{2})}\int_{0}^{1}\,\sum_{k=0}^{+\infty} 
\Big\{\frac{\overline{m}(2\overline{m})^{-k}}{k!\Gamma(\frac{d}{2}-k)}\frac{(-\mbox{ln}\,w)^{d/2-1+k}}{e^{\overline{m}(1-r)}\mp w}+\\
\nonu
&&\frac{(2\overline{m})^{-k}}{k!\Gamma(\frac{d}{2}-k)}\frac{(-\mbox{ln}\;w)^{d/2+k}}{e^{\overline{m}(1-r)}\mp w}
\Big\}.
\er 

In the case of even $d$, we make an important observation. Due to the appearence of the function $\Gamma(d/2-k) (k=0,1,...)$ in the denominator of each term of the series expansion, this series truncates, then containing a finite number of terms up to $k=n/2-1$.

Let us consider thermodynamics in $d=2$ space dimensions. It is fairly simple to obtain a closed expression for $g_{2}$. Setting $d=2$ and  substituting $w=e^{\overline{m}-(x^{2}+\overline{m}^{2})^{1/2}}$ in relation \rf{func} or \rf{sum} one gets 
\br
g_{2}(\overline{m},r) &=& \int_{0}^{1}dw\,\frac{\overline{m}-\mbox{ln}\,w}{\exp[(x^{2}+\overline{m}^{2})^{1/2}-r\overline{m}]\mp1}\\
\nonu
\lab{2dim}
&=& \pm Li_{2}(\pm e^{\overline{m}(r-1)})\mp \overline{m}\,\mbox{ln}(1\mp e^{\overline{m}(r-1)}),
\er
where $Li_{2}$ is the dilogarithm function; see Appendix \ref{appa} for a brief summary of the necessary relations concerning polylogarithms. To derive further thermodynamics it is customary to introduce a parameter $z$ defined by the relationship
\br
z &\equiv & e^{\overline{m}(r-1)},
\er 
the parameter $z$ is generally referred as the fugacity of the system.
 
Then, the density of a relativistic ideal quantum gas in $2D$ (aside from a spin factor) is expressible in the closed and exact form

\br
\rho = \frac{T^{2}}{2\pi}\; \mbox{Sign}(\zeta)\; [Li_{2}(\zeta)+\overline{m} Li_{1}(\zeta)],\;\;\;\; \zeta = \left\{\begin{array}{ll}
z  &\;\;\mbox{if Bose} \\
-z &\;\; \mbox {if Fermi} 
\end{array}
\right.
\lab{rho}
\er
For the Bose gas, the argument $\zeta$ takes on values in the interval $0\leq \zeta \leq 1$. For the Fermi gas, it is $-\infty <\zeta \leq 0$. The relation \rf{rho} is an exact relation valid at any real $\zeta$ ($\zeta \leq 1$), i.e., any temperature \footnote{At temperatures larger than the mass of the particles, QFT requires the inclusion of particle-antiparticle pair production. Therefore the high temperature limit of \rf{rho} is not relevant in realistic physical systems. Thermodynamics of an ultrarelativistic ideal Bose gas is considered in the second paper of references \ct{haber}}.

In the domain of $\zeta$ in which we may approximate $Li_{2}(\zeta)\approx Li_{1}(\zeta)$ (see Fig. below), one can write \rf{rho} as 
\br
\rho = \frac{T^{2}}{2\pi}\; \mbox{Sign}(\zeta)\; [1+\overline{m}] Li_{1}(\zeta),\;\;\;\; \zeta = \left\{\begin{array}{ll}
z  &\;\;\mbox{if Bose}, \\
-z &\;\; \mbox {if Fermi} 
\end{array}
\right.
\lab{rho1}
\er

\setlength{\unitlength}{0.240900pt}
\ifx\plotpoint\undefined\newsavebox{\plotpoint}\fi
\sbox{\plotpoint}{\rule[-0.200pt]{0.400pt}{0.400pt}}%
\begin{picture}(1500,900)(0,0)
\font\gnuplot=cmr10 at 10pt
\gnuplot
\sbox{\plotpoint}{\rule[-0.200pt]{0.400pt}{0.400pt}}%
\put(176.0,473.0){\rule[-0.200pt]{303.534pt}{0.400pt}}
\put(806.0,68.0){\rule[-0.200pt]{0.400pt}{194.888pt}}
\put(176.0,68.0){\rule[-0.200pt]{4.818pt}{0.400pt}}
\put(154,68){\makebox(0,0)[r]{-1}}
\put(1416.0,68.0){\rule[-0.200pt]{4.818pt}{0.400pt}}
\put(176.0,149.0){\rule[-0.200pt]{4.818pt}{0.400pt}}
\put(154,149){\makebox(0,0)[r]{-0.8}}
\put(1416.0,149.0){\rule[-0.200pt]{4.818pt}{0.400pt}}
\put(176.0,230.0){\rule[-0.200pt]{4.818pt}{0.400pt}}
\put(154,230){\makebox(0,0)[r]{-0.6}}
\put(1416.0,230.0){\rule[-0.200pt]{4.818pt}{0.400pt}}
\put(176.0,311.0){\rule[-0.200pt]{4.818pt}{0.400pt}}
\put(154,311){\makebox(0,0)[r]{-0.4}}
\put(1416.0,311.0){\rule[-0.200pt]{4.818pt}{0.400pt}}
\put(176.0,392.0){\rule[-0.200pt]{4.818pt}{0.400pt}}
\put(154,392){\makebox(0,0)[r]{-0.2}}
\put(1416.0,392.0){\rule[-0.200pt]{4.818pt}{0.400pt}}
\put(176.0,473.0){\rule[-0.200pt]{4.818pt}{0.400pt}}
\put(154,473){\makebox(0,0)[r]{0}}
\put(1416.0,473.0){\rule[-0.200pt]{4.818pt}{0.400pt}}
\put(176.0,553.0){\rule[-0.200pt]{4.818pt}{0.400pt}}
\put(154,553){\makebox(0,0)[r]{0.2}}
\put(1416.0,553.0){\rule[-0.200pt]{4.818pt}{0.400pt}}
\put(176.0,634.0){\rule[-0.200pt]{4.818pt}{0.400pt}}
\put(154,634){\makebox(0,0)[r]{0.4}}
\put(1416.0,634.0){\rule[-0.200pt]{4.818pt}{0.400pt}}
\put(176.0,715.0){\rule[-0.200pt]{4.818pt}{0.400pt}}
\put(154,715){\makebox(0,0)[r]{0.6}}
\put(1416.0,715.0){\rule[-0.200pt]{4.818pt}{0.400pt}}
\put(176.0,796.0){\rule[-0.200pt]{4.818pt}{0.400pt}}
\put(154,796){\makebox(0,0)[r]{0.8}}
\put(1416.0,796.0){\rule[-0.200pt]{4.818pt}{0.400pt}}
\put(176.0,877.0){\rule[-0.200pt]{4.818pt}{0.400pt}}
\put(154,877){\makebox(0,0)[r]{1}}
\put(1416.0,877.0){\rule[-0.200pt]{4.818pt}{0.400pt}}
\put(176.0,68.0){\rule[-0.200pt]{0.400pt}{4.818pt}}
\put(176,23){\makebox(0,0){-1}}
\put(176.0,857.0){\rule[-0.200pt]{0.400pt}{4.818pt}}
\put(302.0,68.0){\rule[-0.200pt]{0.400pt}{4.818pt}}
\put(302,23){\makebox(0,0){-0.8}}
\put(302.0,857.0){\rule[-0.200pt]{0.400pt}{4.818pt}}
\put(428.0,68.0){\rule[-0.200pt]{0.400pt}{4.818pt}}
\put(428,23){\makebox(0,0){-0.6}}
\put(428.0,857.0){\rule[-0.200pt]{0.400pt}{4.818pt}}
\put(554.0,68.0){\rule[-0.200pt]{0.400pt}{4.818pt}}
\put(554,23){\makebox(0,0){-0.4}}
\put(554.0,857.0){\rule[-0.200pt]{0.400pt}{4.818pt}}
\put(680.0,68.0){\rule[-0.200pt]{0.400pt}{4.818pt}}
\put(680,23){\makebox(0,0){-0.2}}
\put(680.0,857.0){\rule[-0.200pt]{0.400pt}{4.818pt}}
\put(806.0,68.0){\rule[-0.200pt]{0.400pt}{4.818pt}}
\put(806,23){\makebox(0,0){0}}
\put(806.0,857.0){\rule[-0.200pt]{0.400pt}{4.818pt}}
\put(932.0,68.0){\rule[-0.200pt]{0.400pt}{4.818pt}}
\put(932,23){\makebox(0,0){0.2}}
\put(932.0,857.0){\rule[-0.200pt]{0.400pt}{4.818pt}}
\put(1058.0,68.0){\rule[-0.200pt]{0.400pt}{4.818pt}}
\put(1058,23){\makebox(0,0){0.4}}
\put(1058.0,857.0){\rule[-0.200pt]{0.400pt}{4.818pt}}
\put(1184.0,68.0){\rule[-0.200pt]{0.400pt}{4.818pt}}
\put(1184,23){\makebox(0,0){0.6}}
\put(1184.0,857.0){\rule[-0.200pt]{0.400pt}{4.818pt}}
\put(1310.0,68.0){\rule[-0.200pt]{0.400pt}{4.818pt}}
\put(1310,23){\makebox(0,0){0.8}}
\put(1310.0,857.0){\rule[-0.200pt]{0.400pt}{4.818pt}}
\put(1436.0,68.0){\rule[-0.200pt]{0.400pt}{4.818pt}}
\put(1436,23){\makebox(0,0){1}}
\put(1436.0,857.0){\rule[-0.200pt]{0.400pt}{4.818pt}}
\put(176.0,68.0){\rule[-0.200pt]{303.534pt}{0.400pt}}
\put(1436.0,68.0){\rule[-0.200pt]{0.400pt}{194.888pt}}
\put(176.0,877.0){\rule[-0.200pt]{303.534pt}{0.400pt}}
\put(176.0,68.0){\rule[-0.200pt]{0.400pt}{194.888pt}}
\put(1306,812){\makebox(0,0)[r]{Li1(x)/10}}
\put(1328.0,812.0){\rule[-0.200pt]{15.899pt}{0.400pt}}
\put(176,444){\usebox{\plotpoint}}
\put(176,443.67){\rule{3.132pt}{0.400pt}}
\multiput(176.00,443.17)(6.500,1.000){2}{\rule{1.566pt}{0.400pt}}
\put(201,444.67){\rule{3.132pt}{0.400pt}}
\multiput(201.00,444.17)(6.500,1.000){2}{\rule{1.566pt}{0.400pt}}
\put(189.0,445.0){\rule[-0.200pt]{2.891pt}{0.400pt}}
\put(227,445.67){\rule{3.132pt}{0.400pt}}
\multiput(227.00,445.17)(6.500,1.000){2}{\rule{1.566pt}{0.400pt}}
\put(214.0,446.0){\rule[-0.200pt]{3.132pt}{0.400pt}}
\put(265,446.67){\rule{3.132pt}{0.400pt}}
\multiput(265.00,446.17)(6.500,1.000){2}{\rule{1.566pt}{0.400pt}}
\put(240.0,447.0){\rule[-0.200pt]{6.022pt}{0.400pt}}
\put(291,447.67){\rule{2.891pt}{0.400pt}}
\multiput(291.00,447.17)(6.000,1.000){2}{\rule{1.445pt}{0.400pt}}
\put(278.0,448.0){\rule[-0.200pt]{3.132pt}{0.400pt}}
\put(316,448.67){\rule{3.132pt}{0.400pt}}
\multiput(316.00,448.17)(6.500,1.000){2}{\rule{1.566pt}{0.400pt}}
\put(303.0,449.0){\rule[-0.200pt]{3.132pt}{0.400pt}}
\put(341,449.67){\rule{3.132pt}{0.400pt}}
\multiput(341.00,449.17)(6.500,1.000){2}{\rule{1.566pt}{0.400pt}}
\put(329.0,450.0){\rule[-0.200pt]{2.891pt}{0.400pt}}
\put(367,450.67){\rule{3.132pt}{0.400pt}}
\multiput(367.00,450.17)(6.500,1.000){2}{\rule{1.566pt}{0.400pt}}
\put(354.0,451.0){\rule[-0.200pt]{3.132pt}{0.400pt}}
\put(392,451.67){\rule{3.132pt}{0.400pt}}
\multiput(392.00,451.17)(6.500,1.000){2}{\rule{1.566pt}{0.400pt}}
\put(380.0,452.0){\rule[-0.200pt]{2.891pt}{0.400pt}}
\put(418,452.67){\rule{3.132pt}{0.400pt}}
\multiput(418.00,452.17)(6.500,1.000){2}{\rule{1.566pt}{0.400pt}}
\put(405.0,453.0){\rule[-0.200pt]{3.132pt}{0.400pt}}
\put(443,453.67){\rule{3.132pt}{0.400pt}}
\multiput(443.00,453.17)(6.500,1.000){2}{\rule{1.566pt}{0.400pt}}
\put(431.0,454.0){\rule[-0.200pt]{2.891pt}{0.400pt}}
\put(469,454.67){\rule{2.891pt}{0.400pt}}
\multiput(469.00,454.17)(6.000,1.000){2}{\rule{1.445pt}{0.400pt}}
\put(456.0,455.0){\rule[-0.200pt]{3.132pt}{0.400pt}}
\put(494,455.67){\rule{3.132pt}{0.400pt}}
\multiput(494.00,455.17)(6.500,1.000){2}{\rule{1.566pt}{0.400pt}}
\put(481.0,456.0){\rule[-0.200pt]{3.132pt}{0.400pt}}
\put(520,456.67){\rule{2.891pt}{0.400pt}}
\multiput(520.00,456.17)(6.000,1.000){2}{\rule{1.445pt}{0.400pt}}
\put(507.0,457.0){\rule[-0.200pt]{3.132pt}{0.400pt}}
\put(545,457.67){\rule{3.132pt}{0.400pt}}
\multiput(545.00,457.17)(6.500,1.000){2}{\rule{1.566pt}{0.400pt}}
\put(558,458.67){\rule{3.132pt}{0.400pt}}
\multiput(558.00,458.17)(6.500,1.000){2}{\rule{1.566pt}{0.400pt}}
\put(532.0,458.0){\rule[-0.200pt]{3.132pt}{0.400pt}}
\put(583,459.67){\rule{3.132pt}{0.400pt}}
\multiput(583.00,459.17)(6.500,1.000){2}{\rule{1.566pt}{0.400pt}}
\put(571.0,460.0){\rule[-0.200pt]{2.891pt}{0.400pt}}
\put(609,460.67){\rule{2.891pt}{0.400pt}}
\multiput(609.00,460.17)(6.000,1.000){2}{\rule{1.445pt}{0.400pt}}
\put(621,461.67){\rule{3.132pt}{0.400pt}}
\multiput(621.00,461.17)(6.500,1.000){2}{\rule{1.566pt}{0.400pt}}
\put(596.0,461.0){\rule[-0.200pt]{3.132pt}{0.400pt}}
\put(647,462.67){\rule{3.132pt}{0.400pt}}
\multiput(647.00,462.17)(6.500,1.000){2}{\rule{1.566pt}{0.400pt}}
\put(660,463.67){\rule{2.891pt}{0.400pt}}
\multiput(660.00,463.17)(6.000,1.000){2}{\rule{1.445pt}{0.400pt}}
\put(634.0,463.0){\rule[-0.200pt]{3.132pt}{0.400pt}}
\put(685,464.67){\rule{3.132pt}{0.400pt}}
\multiput(685.00,464.17)(6.500,1.000){2}{\rule{1.566pt}{0.400pt}}
\put(698,465.67){\rule{3.132pt}{0.400pt}}
\multiput(698.00,465.17)(6.500,1.000){2}{\rule{1.566pt}{0.400pt}}
\put(711,466.67){\rule{2.891pt}{0.400pt}}
\multiput(711.00,466.17)(6.000,1.000){2}{\rule{1.445pt}{0.400pt}}
\put(672.0,465.0){\rule[-0.200pt]{3.132pt}{0.400pt}}
\put(736,467.67){\rule{3.132pt}{0.400pt}}
\multiput(736.00,467.17)(6.500,1.000){2}{\rule{1.566pt}{0.400pt}}
\put(749,468.67){\rule{2.891pt}{0.400pt}}
\multiput(749.00,468.17)(6.000,1.000){2}{\rule{1.445pt}{0.400pt}}
\put(761,469.67){\rule{3.132pt}{0.400pt}}
\multiput(761.00,469.17)(6.500,1.000){2}{\rule{1.566pt}{0.400pt}}
\put(723.0,468.0){\rule[-0.200pt]{3.132pt}{0.400pt}}
\put(787,470.67){\rule{3.132pt}{0.400pt}}
\multiput(787.00,470.17)(6.500,1.000){2}{\rule{1.566pt}{0.400pt}}
\put(800,471.67){\rule{2.891pt}{0.400pt}}
\multiput(800.00,471.17)(6.000,1.000){2}{\rule{1.445pt}{0.400pt}}
\put(812,472.67){\rule{3.132pt}{0.400pt}}
\multiput(812.00,472.17)(6.500,1.000){2}{\rule{1.566pt}{0.400pt}}
\put(825,473.67){\rule{3.132pt}{0.400pt}}
\multiput(825.00,473.17)(6.500,1.000){2}{\rule{1.566pt}{0.400pt}}
\put(774.0,471.0){\rule[-0.200pt]{3.132pt}{0.400pt}}
\put(851,474.67){\rule{2.891pt}{0.400pt}}
\multiput(851.00,474.17)(6.000,1.000){2}{\rule{1.445pt}{0.400pt}}
\put(863,475.67){\rule{3.132pt}{0.400pt}}
\multiput(863.00,475.17)(6.500,1.000){2}{\rule{1.566pt}{0.400pt}}
\put(876,476.67){\rule{3.132pt}{0.400pt}}
\multiput(876.00,476.17)(6.500,1.000){2}{\rule{1.566pt}{0.400pt}}
\put(889,477.67){\rule{2.891pt}{0.400pt}}
\multiput(889.00,477.17)(6.000,1.000){2}{\rule{1.445pt}{0.400pt}}
\put(901,478.67){\rule{3.132pt}{0.400pt}}
\multiput(901.00,478.17)(6.500,1.000){2}{\rule{1.566pt}{0.400pt}}
\put(914,479.67){\rule{3.132pt}{0.400pt}}
\multiput(914.00,479.17)(6.500,1.000){2}{\rule{1.566pt}{0.400pt}}
\put(927,480.67){\rule{3.132pt}{0.400pt}}
\multiput(927.00,480.17)(6.500,1.000){2}{\rule{1.566pt}{0.400pt}}
\put(940,481.67){\rule{2.891pt}{0.400pt}}
\multiput(940.00,481.17)(6.000,1.000){2}{\rule{1.445pt}{0.400pt}}
\put(952,482.67){\rule{3.132pt}{0.400pt}}
\multiput(952.00,482.17)(6.500,1.000){2}{\rule{1.566pt}{0.400pt}}
\put(965,483.67){\rule{3.132pt}{0.400pt}}
\multiput(965.00,483.17)(6.500,1.000){2}{\rule{1.566pt}{0.400pt}}
\put(978,485.17){\rule{2.700pt}{0.400pt}}
\multiput(978.00,484.17)(7.396,2.000){2}{\rule{1.350pt}{0.400pt}}
\put(991,486.67){\rule{2.891pt}{0.400pt}}
\multiput(991.00,486.17)(6.000,1.000){2}{\rule{1.445pt}{0.400pt}}
\put(1003,487.67){\rule{3.132pt}{0.400pt}}
\multiput(1003.00,487.17)(6.500,1.000){2}{\rule{1.566pt}{0.400pt}}
\put(1016,488.67){\rule{3.132pt}{0.400pt}}
\multiput(1016.00,488.17)(6.500,1.000){2}{\rule{1.566pt}{0.400pt}}
\put(1029,489.67){\rule{2.891pt}{0.400pt}}
\multiput(1029.00,489.17)(6.000,1.000){2}{\rule{1.445pt}{0.400pt}}
\put(1041,491.17){\rule{2.700pt}{0.400pt}}
\multiput(1041.00,490.17)(7.396,2.000){2}{\rule{1.350pt}{0.400pt}}
\put(1054,492.67){\rule{3.132pt}{0.400pt}}
\multiput(1054.00,492.17)(6.500,1.000){2}{\rule{1.566pt}{0.400pt}}
\put(1067,494.17){\rule{2.700pt}{0.400pt}}
\multiput(1067.00,493.17)(7.396,2.000){2}{\rule{1.350pt}{0.400pt}}
\put(1080,495.67){\rule{2.891pt}{0.400pt}}
\multiput(1080.00,495.17)(6.000,1.000){2}{\rule{1.445pt}{0.400pt}}
\put(1092,497.17){\rule{2.700pt}{0.400pt}}
\multiput(1092.00,496.17)(7.396,2.000){2}{\rule{1.350pt}{0.400pt}}
\put(1105,498.67){\rule{3.132pt}{0.400pt}}
\multiput(1105.00,498.17)(6.500,1.000){2}{\rule{1.566pt}{0.400pt}}
\put(1118,500.17){\rule{2.700pt}{0.400pt}}
\multiput(1118.00,499.17)(7.396,2.000){2}{\rule{1.350pt}{0.400pt}}
\put(1131,501.67){\rule{2.891pt}{0.400pt}}
\multiput(1131.00,501.17)(6.000,1.000){2}{\rule{1.445pt}{0.400pt}}
\put(1143,503.17){\rule{2.700pt}{0.400pt}}
\multiput(1143.00,502.17)(7.396,2.000){2}{\rule{1.350pt}{0.400pt}}
\put(1156,505.17){\rule{2.700pt}{0.400pt}}
\multiput(1156.00,504.17)(7.396,2.000){2}{\rule{1.350pt}{0.400pt}}
\put(1169,507.17){\rule{2.500pt}{0.400pt}}
\multiput(1169.00,506.17)(6.811,2.000){2}{\rule{1.250pt}{0.400pt}}
\put(1181,509.17){\rule{2.700pt}{0.400pt}}
\multiput(1181.00,508.17)(7.396,2.000){2}{\rule{1.350pt}{0.400pt}}
\put(1194,511.17){\rule{2.700pt}{0.400pt}}
\multiput(1194.00,510.17)(7.396,2.000){2}{\rule{1.350pt}{0.400pt}}
\multiput(1207.00,513.61)(2.695,0.447){3}{\rule{1.833pt}{0.108pt}}
\multiput(1207.00,512.17)(9.195,3.000){2}{\rule{0.917pt}{0.400pt}}
\put(1220,516.17){\rule{2.500pt}{0.400pt}}
\multiput(1220.00,515.17)(6.811,2.000){2}{\rule{1.250pt}{0.400pt}}
\multiput(1232.00,518.61)(2.695,0.447){3}{\rule{1.833pt}{0.108pt}}
\multiput(1232.00,517.17)(9.195,3.000){2}{\rule{0.917pt}{0.400pt}}
\multiput(1245.00,521.61)(2.695,0.447){3}{\rule{1.833pt}{0.108pt}}
\multiput(1245.00,520.17)(9.195,3.000){2}{\rule{0.917pt}{0.400pt}}
\multiput(1258.00,524.61)(2.695,0.447){3}{\rule{1.833pt}{0.108pt}}
\multiput(1258.00,523.17)(9.195,3.000){2}{\rule{0.917pt}{0.400pt}}
\multiput(1271.00,527.61)(2.472,0.447){3}{\rule{1.700pt}{0.108pt}}
\multiput(1271.00,526.17)(8.472,3.000){2}{\rule{0.850pt}{0.400pt}}
\multiput(1283.00,530.61)(2.695,0.447){3}{\rule{1.833pt}{0.108pt}}
\multiput(1283.00,529.17)(9.195,3.000){2}{\rule{0.917pt}{0.400pt}}
\multiput(1296.00,533.60)(1.797,0.468){5}{\rule{1.400pt}{0.113pt}}
\multiput(1296.00,532.17)(10.094,4.000){2}{\rule{0.700pt}{0.400pt}}
\multiput(1309.00,537.60)(1.651,0.468){5}{\rule{1.300pt}{0.113pt}}
\multiput(1309.00,536.17)(9.302,4.000){2}{\rule{0.650pt}{0.400pt}}
\multiput(1321.00,541.59)(1.378,0.477){7}{\rule{1.140pt}{0.115pt}}
\multiput(1321.00,540.17)(10.634,5.000){2}{\rule{0.570pt}{0.400pt}}
\multiput(1334.00,546.59)(1.123,0.482){9}{\rule{0.967pt}{0.116pt}}
\multiput(1334.00,545.17)(10.994,6.000){2}{\rule{0.483pt}{0.400pt}}
\multiput(1347.00,552.59)(1.123,0.482){9}{\rule{0.967pt}{0.116pt}}
\multiput(1347.00,551.17)(10.994,6.000){2}{\rule{0.483pt}{0.400pt}}
\multiput(1360.00,558.59)(0.874,0.485){11}{\rule{0.786pt}{0.117pt}}
\multiput(1360.00,557.17)(10.369,7.000){2}{\rule{0.393pt}{0.400pt}}
\multiput(1372.00,565.59)(0.728,0.489){15}{\rule{0.678pt}{0.118pt}}
\multiput(1372.00,564.17)(11.593,9.000){2}{\rule{0.339pt}{0.400pt}}
\multiput(1385.00,574.58)(0.539,0.492){21}{\rule{0.533pt}{0.119pt}}
\multiput(1385.00,573.17)(11.893,12.000){2}{\rule{0.267pt}{0.400pt}}
\multiput(1398.58,586.00)(0.493,0.616){23}{\rule{0.119pt}{0.592pt}}
\multiput(1397.17,586.00)(13.000,14.771){2}{\rule{0.400pt}{0.296pt}}
\multiput(1411.58,602.00)(0.492,1.186){21}{\rule{0.119pt}{1.033pt}}
\multiput(1410.17,602.00)(12.000,25.855){2}{\rule{0.400pt}{0.517pt}}
\multiput(1423.58,630.00)(0.493,8.506){23}{\rule{0.119pt}{6.715pt}}
\multiput(1422.17,630.00)(13.000,201.062){2}{\rule{0.400pt}{3.358pt}}
\put(838.0,475.0){\rule[-0.200pt]{3.132pt}{0.400pt}}
\put(1306,767){\makebox(0,0)[r]{Li2(x)/10}}
\multiput(1328,767)(20.756,0.000){4}{\usebox{\plotpoint}}
\put(1394,767){\usebox{\plotpoint}}
\put(176,439){\usebox{\plotpoint}}
\put(176.00,439.00){\usebox{\plotpoint}}
\put(196.72,440.00){\usebox{\plotpoint}}
\multiput(201,440)(20.694,1.592){0}{\usebox{\plotpoint}}
\put(217.42,441.26){\usebox{\plotpoint}}
\put(238.15,442.00){\usebox{\plotpoint}}
\multiput(240,442)(20.684,1.724){0}{\usebox{\plotpoint}}
\put(258.87,443.00){\usebox{\plotpoint}}
\multiput(265,443)(20.694,1.592){0}{\usebox{\plotpoint}}
\put(279.58,444.00){\usebox{\plotpoint}}
\put(300.31,444.78){\usebox{\plotpoint}}
\multiput(303,445)(20.694,1.592){0}{\usebox{\plotpoint}}
\put(321.01,446.00){\usebox{\plotpoint}}
\multiput(329,446)(20.684,1.724){0}{\usebox{\plotpoint}}
\put(341.73,447.00){\usebox{\plotpoint}}
\put(362.46,447.65){\usebox{\plotpoint}}
\multiput(367,448)(20.694,1.592){0}{\usebox{\plotpoint}}
\put(383.16,449.00){\usebox{\plotpoint}}
\put(403.88,449.91){\usebox{\plotpoint}}
\multiput(405,450)(20.694,1.592){0}{\usebox{\plotpoint}}
\put(424.60,451.00){\usebox{\plotpoint}}
\multiput(431,451)(20.684,1.724){0}{\usebox{\plotpoint}}
\put(445.30,452.18){\usebox{\plotpoint}}
\put(466.03,453.00){\usebox{\plotpoint}}
\multiput(469,453)(20.684,1.724){0}{\usebox{\plotpoint}}
\put(486.72,454.44){\usebox{\plotpoint}}
\multiput(494,455)(20.756,0.000){0}{\usebox{\plotpoint}}
\put(507.46,455.04){\usebox{\plotpoint}}
\put(528.15,456.68){\usebox{\plotpoint}}
\multiput(532,457)(20.756,0.000){0}{\usebox{\plotpoint}}
\put(548.88,457.30){\usebox{\plotpoint}}
\put(569.57,458.89){\usebox{\plotpoint}}
\multiput(571,459)(20.756,0.000){0}{\usebox{\plotpoint}}
\put(590.30,459.56){\usebox{\plotpoint}}
\multiput(596,460)(20.694,1.592){0}{\usebox{\plotpoint}}
\put(611.00,461.00){\usebox{\plotpoint}}
\put(631.73,461.83){\usebox{\plotpoint}}
\multiput(634,462)(20.694,1.592){0}{\usebox{\plotpoint}}
\put(652.42,463.42){\usebox{\plotpoint}}
\multiput(660,464)(20.756,0.000){0}{\usebox{\plotpoint}}
\put(673.15,464.09){\usebox{\plotpoint}}
\put(693.84,465.68){\usebox{\plotpoint}}
\multiput(698,466)(20.694,1.592){0}{\usebox{\plotpoint}}
\put(714.55,467.00){\usebox{\plotpoint}}
\put(735.27,467.94){\usebox{\plotpoint}}
\multiput(736,468)(20.694,1.592){0}{\usebox{\plotpoint}}
\put(755.96,469.58){\usebox{\plotpoint}}
\multiput(761,470)(20.756,0.000){0}{\usebox{\plotpoint}}
\put(776.69,470.21){\usebox{\plotpoint}}
\put(797.38,471.80){\usebox{\plotpoint}}
\multiput(800,472)(20.684,1.724){0}{\usebox{\plotpoint}}
\put(818.07,473.47){\usebox{\plotpoint}}
\multiput(825,474)(20.694,1.592){0}{\usebox{\plotpoint}}
\put(838.77,475.00){\usebox{\plotpoint}}
\put(859.49,475.71){\usebox{\plotpoint}}
\multiput(863,476)(20.694,1.592){0}{\usebox{\plotpoint}}
\put(880.19,477.32){\usebox{\plotpoint}}
\put(900.87,478.99){\usebox{\plotpoint}}
\multiput(901,479)(20.694,1.592){0}{\usebox{\plotpoint}}
\put(921.57,480.58){\usebox{\plotpoint}}
\multiput(927,481)(20.694,1.592){0}{\usebox{\plotpoint}}
\put(942.26,482.19){\usebox{\plotpoint}}
\put(962.98,483.00){\usebox{\plotpoint}}
\multiput(965,483)(20.694,1.592){0}{\usebox{\plotpoint}}
\put(983.68,484.44){\usebox{\plotpoint}}
\multiput(991,485)(20.684,1.724){0}{\usebox{\plotpoint}}
\put(1004.37,486.11){\usebox{\plotpoint}}
\put(1025.07,487.70){\usebox{\plotpoint}}
\multiput(1029,488)(20.684,1.724){0}{\usebox{\plotpoint}}
\put(1045.76,489.37){\usebox{\plotpoint}}
\put(1066.45,490.96){\usebox{\plotpoint}}
\multiput(1067,491)(20.694,1.592){0}{\usebox{\plotpoint}}
\put(1087.07,493.18){\usebox{\plotpoint}}
\multiput(1092,494)(20.694,1.592){0}{\usebox{\plotpoint}}
\put(1107.71,495.21){\usebox{\plotpoint}}
\put(1128.40,496.80){\usebox{\plotpoint}}
\multiput(1131,497)(20.684,1.724){0}{\usebox{\plotpoint}}
\put(1149.09,498.47){\usebox{\plotpoint}}
\multiput(1156,499)(20.694,1.592){0}{\usebox{\plotpoint}}
\put(1169.78,500.13){\usebox{\plotpoint}}
\put(1190.35,502.72){\usebox{\plotpoint}}
\multiput(1194,503)(20.694,1.592){0}{\usebox{\plotpoint}}
\put(1211.01,504.62){\usebox{\plotpoint}}
\put(1231.62,506.97){\usebox{\plotpoint}}
\multiput(1232,507)(20.694,1.592){0}{\usebox{\plotpoint}}
\put(1252.25,509.12){\usebox{\plotpoint}}
\multiput(1258,510)(20.694,1.592){0}{\usebox{\plotpoint}}
\put(1272.87,511.31){\usebox{\plotpoint}}
\put(1293.46,513.80){\usebox{\plotpoint}}
\multiput(1296,514)(20.514,3.156){0}{\usebox{\plotpoint}}
\put(1314.04,516.42){\usebox{\plotpoint}}
\multiput(1321,517)(20.514,3.156){0}{\usebox{\plotpoint}}
\put(1334.61,519.09){\usebox{\plotpoint}}
\put(1355.12,522.25){\usebox{\plotpoint}}
\multiput(1360,523)(20.473,3.412){0}{\usebox{\plotpoint}}
\put(1375.61,525.56){\usebox{\plotpoint}}
\put(1396.13,528.71){\usebox{\plotpoint}}
\multiput(1398,529)(20.224,4.667){0}{\usebox{\plotpoint}}
\put(1416.44,532.91){\usebox{\plotpoint}}
\multiput(1423,534)(20.224,4.667){0}{\usebox{\plotpoint}}
\put(1436,537){\usebox{\plotpoint}}
\end{picture}\\
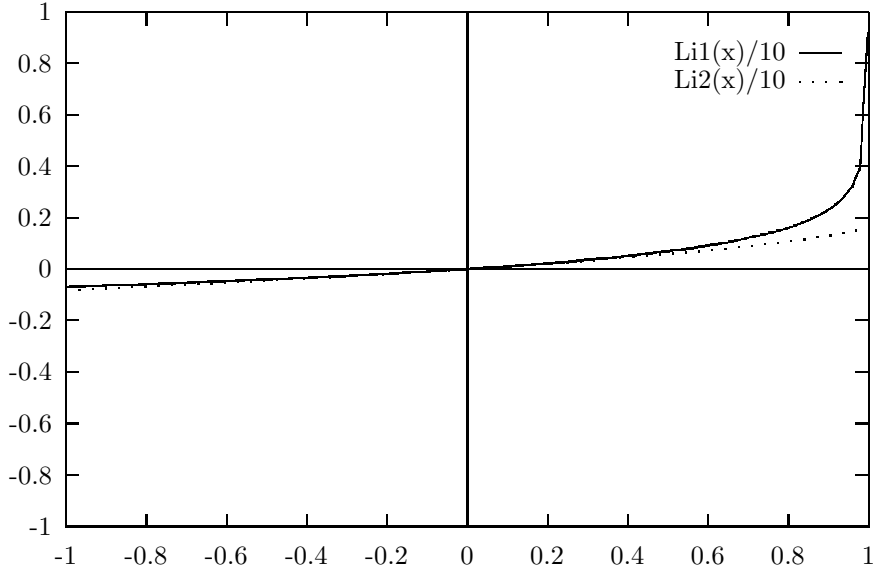
Fig. Observe the domain of $\zeta$ in which we may approximate $Li_{2}(\zeta)\approx Li_{1}(\zeta)$.\\

Then, taking into account \rf{rho1} and the recurrence relation \ref{rec} for $m=1$, we may obtain the basic thermodynamic quantities- pressure P, energy U, and entropy S:

\br
\lab{pres}
\rho^{-1} \beta P&=& \frac{Li_{2}(\zeta)}{Li_{1}(\zeta)},\\
\lab{ener}
\beta\frac{U}{N}&=&\frac{Li_{2}(\zeta)}{Li_{1}(\zeta)},\\
\lab{entro}
\frac{S}{N}&=&2 \frac{Li_{2}(\zeta)}{Li_{1}(\zeta)}-\mbox{ln}|\zeta|+\overline{m}.
\er

Following the same procedure as in \ct{lee2}, we can assume the same reduced density for both type of particles  

\br
\lab{equal}
(\frac{2\pi}{T^{2}})\rho =[1+\overline{m}] Li_{1}(z_{B})=-[1+\overline{m}] Li_{1}(-z_{F}),
\er
where $z_B$ and $z_{F}$ are the fugacities of the Bose and Fermi gases, respectively. Assuming that both type of particles are spinless and have equal mass, \rf{equal} allows us to consider both gases at the same temperature $T$. Using the Landen's relation \rf{lan2} in \rf{equal} we can deduce that the fugacities are related by the following Euler transformation
\br
\lab{euler} 
 z_{F} =z_{B}/(1-z_{B}).
\er

Applying Landen's relations \ref{lan2} and \ref{lan3} to \rf{pres}-\rf{entro} we may obtain, for example for the energy
\br
\nonu
\beta \frac{U(z_{B})}{N}&=& \frac{Li_{2}(z_{B})}{Li_{1}(z_{B})}\\
&=&\frac{Li_{2}(-z_{B}/(1-z_{B}))+\h [(Li_{1}(-z_{B}/(1-z_{B})))]^2}{Li_{1}(-z_{B}/(1-z_{B}))}.
\er
    
By \rf{euler} we may obtain 
\br
\nonu
\frac{U(z_{B})}{N}&=&\frac{U(z_{F})}{N}+\h Li_{1}(-z_{F})\\
\lab{energy}
&=&\frac{U(z_{F})}{N}-\h \frac{1}{1+\overline{m}}\frac{2\pi \rho}{T}.
\er 

The second term on the right hand side of \rf{energy} is $T$ dependent. It corresponds to the relativistic correction introduced, considering the $Li_{2}(\zeta)$ term in \rf{rho1}, in the energy relationship of both type of gases.

The non-relativistic limit corresponds to $T\ll m$. In this limit the contribution of the first term in \rf{rho} is negligible, i.e., the $\frac{T^{2}}{2\pi} Li_{2}(\zeta)$ function, is smaller than the $\frac{\overline{m}}{2\pi} T^{2} Li_{1}(\zeta)$ term. The term proportional to $\frac{T^{2}}{2\pi} Li_{2}(\zeta)$, thus, becomes a relativistic correction to the expression presented in \ct{lee2}, in which only the monolog term, $ Li_{1}(\zeta)$ appears.

In the limit  $\overline{m}>>1$ (non-relativistic limit) one gets from \rf{energy} the same relationship between the energies of both type of gases as in \ct{lee2}. Since the second term of the right hand side of \rf{energy} in this approximation will not depend on $T$, of course the specific heats must be the same. The same analysis can be performed for the remaining themodynamic quantities. Then, in the  $\overline{m}>>1$ and $Li_{2}(\zeta)\approx Li_{1}(\zeta)$ approximations, the results of Lee \ct{lee2} can be recovered if one considers from the beginning the study of ideal relativistic gases; i.e. ``in $d=2$ the energy and pressure of the Bose gas are equal to those of the Fermi gas shifted by the zero-point constant of the Fermi gas''.     
   
\section{Conclusions}

In this paper we derived the expression for the densities of relativistic ideal Bose and Fermi gases in two dimensions in a closed and exact form making use of polylogarithm functions. Taking into account the approximation $Li_{2}(\zeta)\approx Li_{1}(\zeta)$, valid for some domain of $\zeta$,  we have related the energies of both type of gases, showing that they differ by a temperature $T$ dependent term. We then concluded that essentially both gases are inequivalent even in this approximation.   
    
 We rederived the results of \ct{lee2} making a further approximation, $\overline{m}>>1$, and thus stablished the equivalence between the spinless and equal mass Bose and Fermi gases in the non-relativistic limit.

\section{Acknowledgements}
H.B is thankful to FAPESP (grant number 96/00212-0), and B.M.P. and J.L.T. to CNPq, for partial financial support.

\appendix
\section{The dilog and monolog relationships}

\label{appa}
The following properties of polylogs can be stablished, see \ct{lewin,kirilov} for details. 
We have the recurrence relation  
\br
\label{rec}
Li_{m}(x)\;=\;(x\frac{d}{dx})Li_{m+1}(x).
\er
If $x$ is real number and $x<1$ and $y\;=\;-\frac{x}{1-x}$, then 
\br
\label{lan2}
Li_{1}(x)&=&-Li_{1}(y)\; 
\er
and
\br
\label{lan3}
Li_{2}(x)\;=\;-Li_{2}(y)-\h (Li_{1}(y))^2.
\er

\end{document}